\documentclass[12pt]{article}

\newif\ifpdf
\ifx\pdfoutput\undefined
\pdffalse % we are not running PDFLaTeX
\else
\pdfoutput=1 % we are running PDFLaTeX
\pdftrue
\fi

\ifpdf
\usepackage[pdftex]{graphicx}
\else
\usepackage{graphicx}
\fi

\usepackage{setspace}
\title{
{\normalsize \hfill SPIN-2003/14}\\
\vspace{-1.7cm}
{\normalsize \hfill ITP-UU-03/24}\\
${}$\\ 
Sum over topologies and double-scaling limit in 2D Lorentzian 
quantum gravity
}

\author{
R. Loll\footnote{email: r.loll@phys.uu.nl} \  and 
W. Westra\footnote{email: w.westra@phys.uu.nl} \\
${}$\\
{\small Institute for Theoretical Physics, Utrecht University,}\\
{\small Leuvenlaan 4, NL-3584 CE Utrecht}
}

\begin{document}

\ifpdf
\DeclareGraphicsExtensions{.pdf, .jpg, .tif}
\else
\DeclareGraphicsExtensions{.eps, .jpg}
\fi

\maketitle
\abstract{
We construct a combined non-perturbative path integral 
over geometries and topologies for two-dimensional
Lorentzian quantum gravity. 
The Lorentzian structure is used in an essential way 
to exclude geometries with unacceptably large causality 
violations. The remaining
sum can be performed analytically and possesses a 
unique and well-defined double-scaling limit, a property
which has eluded similar models of Euclidean quantum
gravity in the past. 
}

\section{Summing over topologies?}
A central question that arises in the construction of a theory of 
quantum gravity is that of the fundamental, microscopic degrees of 
freedom whose dynamics the theory should describe. 
The idea that the information contained in the metric  field tensor
$g_{\mu\nu}$ may not constitute an
adequate description of the geometric properties of space-time
at the very shortest scales goes back all the way to Riemann
himself \cite{riemann}. More recently this has led to the suggestion
that at the Planck scale also the topological degrees of freedom
of space-time should become excited. Wheeler is usually credited
with coining the notion of a space-time foam \cite{wheeler},
according to which space-time is a smooth, classical
object macroscopically, well described by general relativity, but at 
the Planck scale presents a scenario of wildly fluctuating 
geometry and topology. 

In the context of the gravitational path integral, this has inspired
an extension of the customary integral over all metrics (modulo
diffeomorphisms) by an additional sum over space-time
topologies, namely,
\begin{equation}
Z(\kappa,\lambda)=\sum_{\rm topol.} \int D[g_{\mu\nu}]{\rm e}^{iS[g_{\mu\nu}]},
\label{pi}
\end{equation}
where the square brackets denote diffeomorphism equivalence
classes of metrics and where $S$ is the gravitational action.
We will take the action to include a cosmological term,
\begin{equation}
S=\int d^dx \sqrt{|\det g|} (\kappa R-\lambda),
\label{action}
\end{equation}
with $\kappa$ and $\lambda$ denoting the inverse gravitational
coupling constant and the cosmological constant. 
In space-time dimension $d=4$, given the well-known difficulties of 
defining a path integral over the metric degrees of freedom alone, 
it may not come as a surprise that very little progress has been
made in giving a well-defined mathematical and physical
meaning to (\ref{pi}). Previous semi-classical treatments of topology 
change, as for example in discussions of the effect of
baby universes on effective coupling constants (see \cite{giddings} for
a critical appraisal) are unlikely to be of relevance to the problem,
for a variety of reasons. 

First, since there is currently no direct or indirect evidence for topology 
changes from experiment, the phenomenon -- if realized at all -- must 
take place at the Planck scale or not too far from it, casting
doubt on the applicability of semi-classical methods. Secondly, with
very few exceptions, such investigations
have been made within the path integral for {\it Euclidean} metrics. 
However, in the absence of a Wick rotation for theories with a dynamical 
metric, the Euclidean theory has no obvious relation with the physical,
Lorentzian theory. Moreover, the Euclidean path integral seems to suffer
from incurable divergences due to the presence of the
conformal mode \cite{conformal}. Lastly, and most importantly, once
topology change is permitted, topology-changing contributions
dominate the path integral completely, since the number of distinct 
geometries at a fixed space-time volume $V$ grows {\it super-exponentially} 
with $V$. This entropic effect is truly non-perturbative and cannot be seen 
in a semi-classical treatment. It implies that arguments for a dynamical
suppression of topology changes which are based on an evaluation of
their semi-classical action are largely irrelevant. 

Yet more worrying for the proponents of a ``sum over
topologies" should be the fact that the analogous problem is unsolved
even in dimension $d<4$.\footnote{Quantum gravities in dimension 2 and 3 serve
as useful models for diffeomorphism-invariant theories of dynamical
geometry. Their metric configuration spaces and dynamics are much simplified 
in comparison with the physical, four-dimensional theory.} 
Again, this can be traced to the super-exponential growth of the
number of geometries with their volume, which renders the path integral
badly divergent. 

In space-time dimension two, which we will focus on in the following,
the quantization of pure Euclidean (or Liouville) gravity for {\it fixed} topology 
is well understood in analytic terms \cite{2deuclid}. 
The sum over topologies is turned into
a sum over a single parameter $g\geq 0$, the genus (number of
handles or holes) of the two-dimensional space-time. The Euclidean analogue of
the path integral (\ref{pi}) for $d=2$ has been the object of intense study in the past,
since it is an example of a non-perturbative sum over world sheets of a
bosonic string (in a zero-dimensional target space) \cite{3papers}.
The problem has been addressed by matrix model methods or, equivalently,
a regularization of the path integral in terms of triangulated, piecewise
flat two-surfaces.
However, it turns out that the topological expansion of (\ref{pi}) in powers of
${\rm e}^{-\kappa}$ (the integrated curvature in 2d is proportional to $g$, up
to an additive constant) is not Borel-summable, because the coefficients
in the series grow factorially with $g$ and are all positive. Attempts to fix
the ensuing non-perturbative ambiguities of the partition function 
in a unique and physically motivated way have so far remained unsuccessful 
\cite{2deuclid}.

\section{Doing it the Lorentzian way}

This leaves us in the rather unsatisfactory situation of not having a
single instance of a quantum-gravitational theory where the sum over
topologies had actually been performed. In the present work, we will
suggest a possible way out of this impasse.
The central idea is to take
seriously the causal nature of space-time, and to perform a 
non-perturbative summation over {\it Lorentzian} geometries. As
regards the sum over topologies, the Lorentzian structure will be used
to quantify how badly causality is violated by individual contributions
to the path integral. We will introduce and solve a model of 2d quantum
gravity which at the regularized level amounts to a sum over 2d 
piecewise linear space-times of any genus
whose causal properties are ``not too bad". For the purposes of this
paper, we will adopt a strictly
quantum-gravitational point of view, in the sense that we will only be
interested in models that do not lead to large-scale causality 
violations. In particular, we do not think that in this context third-quantized 
models, whose Hilbert spaces describe multiply-connected 
spatial geometries, can be interpreted in a physically meaningful
way. This is 
different from situations where the geometries appear as
imbedded quantities, as they do in the case of string theory, where
moreover topology changes of the world sheet are mandatory, and not optional. 

The question is then whether there are any models with topology 
change that produce a quantum space-time foam whose non-trivial 
microstructure leads to a measurable, but not necessarily large 
effect at a macroscopic level. 
The quantum gravity model we are about to construct has exactly this property. 
Although it is a model whose topological fluctuations are associated
with the ``mildest" type of causality violation imaginable in two
dimensions,\footnote{However, it should be kept in mind that all 
topology changes
in 2 and 3d which do not involve universe creation or annihilation
are ``bad'' according to the classification of Dowker and collaborators 
\cite{dowker}.} 
it is already at the limit of what is
acceptable as a space-time foam. Namely, we will show that for a
sufficiently large value of the renormalized gravitational coupling
constant, the effects of topology change become overwhelming
and the system enters a phase of ``handle condensation".

In order to perform the sum (\ref{pi}) non-perturbatively, we will adopt
a Lorentzian version of the regularized sum over piecewise flat
2d space-times. For fixed topology $[0,1]\times S^1$, this model
is exactly soluble and leads to a 2d quantum gravity theory 
inequivalent to Liouville quantum gravity \cite{lor2d}, with a
well-defined Wick rotation and
without a $c=1$ barrier \cite{barrier}. The difference
can be traced to the absence (in the Lorentzian case) of branching
``baby universes", which are incompatible with causality.\footnote{By
this we mean baby universes that do not return to the ``mother
universe", and therefore do not change the {\it space-time} topology.
We will not consider such configurations in the present work.} This method of
``Lorentzian dynamical triangulations" has also been applied
successfully in dimension three \cite{lor3d}, leading to a well-behaved 
quantum ground state, which unlike in the Euclidean theory does not
degenerate into a lower-dimensional polymer as a consequence of a 
dominance of the conformal mode. 
\begin{figure}[tbp]
\centerline{\scalebox{0.6}{\rotatebox{0}
{\includegraphics{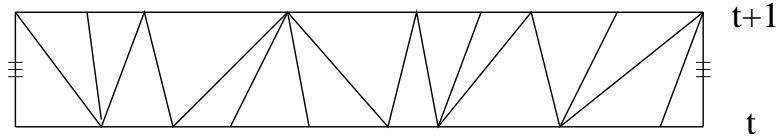}}}}
\caption[onestrip]{A strip $[t,t+1]$ of a 2d Lorentzian triangulated
space-time. The ends of the strip should be identified as indicated,
leading to a compact spatial geometry $S^1$.}
%\vspace{0.1cm}
\label{onestrip}
\end{figure}

Recall that any 1+1 dimensional Lorentzian space-time contributing
to the regularized path integral is given by a sequence of strips of
height $\Delta t=1$, where each strip in turn is a random sequence of $N$
Minkowskian up- and down-triangles (Fig.\ref{onestrip}), 
each with two time-like and one
space-like edge of length-squared $\pm a^2$ \cite{lor2d}. We will now
generalize these to a class of Lorentzian geometries with holes, where
the holes have minimal time duration $\Delta t=1$. Although this time
interval goes to zero in the continuum limit $a\rightarrow 0$, their
effect is not necessarily negligible, since the triangle density goes to
infinity in the limit. 
\begin{figure}[tbp]
\centerline{\scalebox{0.4}{\rotatebox{0}
{\includegraphics{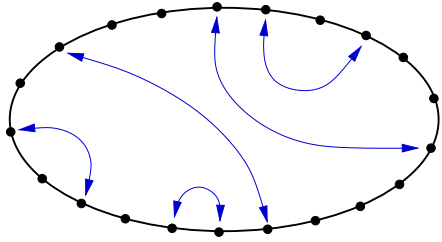}}}}
\caption[arch]{Horizontal section through a strip $[t,t+1]$ of triangles,
with time-like edges appearing as dots. The arrows indicate where the
circle should be glued and cut to obtain (in this case six) disconnected
cylinder components.}
%\vspace{0.1cm}
\label{arch}
\end{figure}
The way in which we create triangulations with holes in a strip $[t,t+1]$ is as
follows. Suppose the geometry has been built up to integer time $t$,
possibly with holes. The spatial geometry at time $t$ is a closed circle
consisting of $l_t$ space-like edges. Now, glue on another strip 
with matching ``in-geometry" of length $l_t$ and some ``out-geometry"
of length $l_{t+1}$ (giving rise to a total discrete strip volume of $N_t=l_t+l_{t+1}$
triangles). Next, glue an even number of the $N_t$ time-like edges 
in the strip pairwise to each other, according to an arrow diagram (Fig.\ref{arch}),
and then cut open the geometry at each of these edges, perpendicular
to the direction in which they were glued together. This will result in
a space-time geometry consisting of several cylindrical components
between $t$ and $t+1$. In order to obtain back a spatial circle at
time $t+1$, the cylinders must be cut open at some of their vertices
at $t+1$ and their spatial boundaries be aligned in some order to
form again a single $S^1$. In this way, one has constructed a 
strip geometry with some number $g_t$ of holes. 

It is straightforward to show that if arbitrary regluings at time $t+1$
are allowed, the number of possible geometries at a given strip volume
scales factorially with $N_t$, just as in the Euclidean case. 
However,
if one looks at the causal properties of the resulting space-times,
most of them turn out to be extremely ill-behaved, in the sense that
even a single hole in the entire space-time will lead to a global
rearrangement of parts of a light front after passing the hole, even
if it exists only for an infinitesimal time $\Delta t$. Fortunately, there is
a subclass of geometries for which this does not happen, which
are those where the cylinders are reglued {\it without}
any intermediate relative twisting or rearrangement of the order of the 
components during the time interval when they are disconnected. 

The effect is most easily illustrated by the case of two components.
In a ``regluing without twist"  the two saddle points\footnote{Note
in passing that it is not clear a priori how to account for
the curvature singularities at the saddle points in the Lorentzian
action, and how to treat them in the Wick rotation, see \cite{lousor}
for a related discussion. We will simply use the standard Regge
prescription in terms of deficit angles in the Wick-rotated action.}
$p_t$ and $p_{t+1}$ at $t$ and $t+1$
where the hole appears and vanishes are connected in each of
the two cylinders by a time-like link, which implies that they are
nearest lattice neighbours in either of the components. If this is
not the case, e.g. if one of the cylinders is twisted before regluing,
then $p_t$ and $p_{t+1}$ will not appear as nearest neighbours
in that component, but $p_{t+1}$ will have a relative shift
$\Delta l$ along the spatial direction. The resulting space-time 
geometry will have the property that a light beam of macroscopic
width that passes by the hole will be split into two parts which
will emerge with a relative separation of $\Delta l$ after the hole 
disappears!
By comparison, the only effect of the hole in the ``untwisted" case
is that a small fraction of the light beam will be scattered into the
far-away part of the space-time to which the hole connects during
its infinitesimal life time (this effect is of course 
also present in the twisted case).\footnote{A detailed geometric analysis 
can be found in \cite{prep}.} Since we find it difficult to envisage how
a quantum geometry with anything near a macroscopic causal
structure could emerge from a superposition of such ill-behaved 
manifolds, our sum over topologies will contain only geometries
with ``untwisted" holes.

\section{Discrete solution and double-scaling limit}

\begin{figure}[tbp]
%\vspace{0.5cm}
\centerline{\scalebox{0.9}{\rotatebox{0}
{\includegraphics{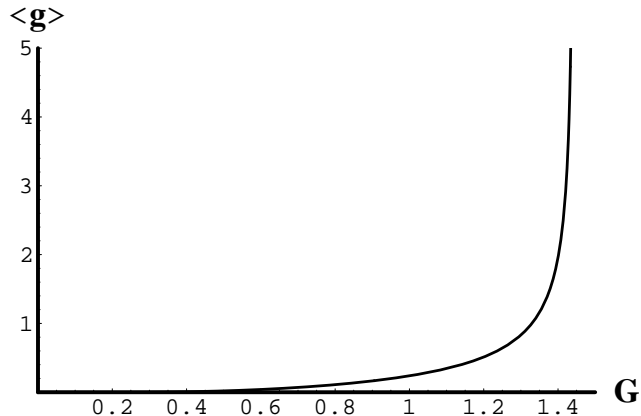}}}}
\caption[genplot]{The average number $\langle g\rangle$ of holes as a function
of Newton's constant $G$ in the one-strip model described by the
partition function (\ref{zdone}). }
%\vspace{0.1cm}
\label{genplot}
\end{figure}

To illustrate that the causality constraints imposed above do lead to
a well-defined and soluble model, 
we will now solve the combinatorics for a single strip $\Delta t=1$ and 
look for a scaling behaviour of the two coupling constants that leads
to a non-trivial continuum
limit. The partition function after Wick-rotating is 
\begin{equation}
Z(\lambda,\kappa)=\sum_{l_{in}}\sum_{l_{out}} 
{\rm e}^{-\lambda (l_{in}+l_{out})}
\sum_{T|_{l_{in},l_{out}} }{\rm e}^{-\kappa g(T)}, 
\label{part}
\end{equation}
with a sum over the initial and final
boundary geometries of length $l_{in}$ and $l_{out}$, and a sum
over triangulations $T$ of a strip with these boundaries.
For a given triangulated strip of volume $N=l_{in}+l_{out}$, the 
counting of geometries with holes according to the procedure
of the previous section involves a counting of diagrams like in
Fig.\ref{arch} with $N$ vertices and $g$ arrows. Expression (\ref{part})
can be rearranged,
\begin{equation}
Z(\lambda,\kappa)= \frac{1}{2} \sum_{N=0}^\infty\ \sum_{g=0}^{[N/2]}\  
\biggl({N\atop 2g}\biggr)
\ \frac{(2g)!}{g! (g+1)!}\  {\rm e}^{-2\kappa g} {\rm e}^{-(\lambda -\log 2)N},
\label{z1}
\end{equation}
after which the sums can be performed explicitly, leading
to
\begin{equation}
Z(\lambda,\kappa)=\frac{1}{2 (1-{\rm e}^{-(\lambda-\log 2)})}\ \frac{1-\sqrt{1-4 z}}{2z},
\label{partz}
\end{equation}
where the second term depends only on the combination
\begin{equation}
z:={\rm e}^{-2\kappa}({\rm e}^{\lambda -\log 2}-1)^{-2}.
\label{both}
\end{equation}
An infinite-volume limit is obtained by tuning the bare cosmological
coupling $\lambda$ to $\log 2$ {\it from above}\footnote{Note that this gives
rise to a non-negative renormalized cosmological constant; our approach
naturally leads to a de-Sitter-like behaviour.},
as in standard Lorentzian quantum gravity \cite{lor2d},
\begin{equation}
\lambda =\lambda^{crit}+a^2\Lambda +O(a^3)\equiv
 \log 2 +a^2\Lambda +O(a^3),
\label{cosren}
\end{equation}
where $\Lambda$ denotes the renormalized, dimensionful cosmological
constant, as the geodesic cutoff $a\rightarrow 0$. As can be seen from 
eq.(\ref{partz}), this is only consistent if simultaneously also the inverse
Newton constant $\kappa$ is renormalized. Such a double-scaling
limit is obtained by fixing $z$ to a constant, $z=c<1/4$,
and defining a renormalized coupling $\rm K$ by
\begin{equation}
{\rm K} =\kappa -2 \log \frac{1}{a\sqrt{\Lambda}} +O(a),\;\;\;\;
{\rm K}:=\frac{1}{2} \log \frac{1}{c}.
\label{kapren}
\end{equation}
Substituting these expansions into the expression for the strip
partition function (\ref{partz}), 
a straightforward computation yields the renormalized partition
function in terms of $\Lambda$ and the gravitational coupling $G=1/$K,
\begin{equation}
Z^R(\Lambda,G)=
\frac{ {\rm e}^{2/G}}{4\Lambda } 
\biggl( 1-\sqrt{1-4\, {\rm e}^{-2/G}}
\biggr).
\label{zdone}
\end{equation}
In the continuum theory, one expects $\Lambda$ to set the global scale 
because of
$\langle V\rangle =\frac{1}{\Lambda}$ for the expectation value
of the space-time volume. On the other hand, the strength of the 
gravitational coupling governs the average number $\langle g
\rangle$ of holes per slice, which is proportional to the fraction
of a lightbeam that will be scattered by holes in a non-local and
causality-violating manner \cite{prep}. As is illustrated by 
Fig.\ref{genplot}, for $G=0$ there are no holes at all. Their number
increases for $G>0$, first slowly and then rapidly, until it
diverges at the maximum value $G=2/\log 4$, at which point the
system undergoes a transition to a phase of ``condensed handles".

What we have found therefore is
an example of a gravity-inspired statistical model with a well-defined double-scaling
limit. As in previous work on non-perturbative gravitational path
integrals, the Lorentzian structure of the individual geometries has
played a crucial role in the construction. In forthcoming work \cite{lwz}
we will investigate what happens when one keeps the boundaries of the
space-time strip fixed instead of summing over them, as presented here.
This more complicated model needs to be solved in order to determine
the Hamiltonian and the full propagator of 2D Lorentzian quantum gravity
with holes. Interestingly, it turns out that the inclusion of the boundaries
leads to a different scaling behaviour of Newton's constant. This also implies
a different behaviour for the number of holes: unlike in the strip model,
there is no condensation of handles, and the number of holes per strip
stays infinitesimal. Unlike in the
pure Lorentzian theory without holes therefore, the scaling of the ``bulk" 
couplings and
the bulk partition function cannot be deduced from solving the
simpler strip model with summed-over boundaries. This may teach us
an important lesson for higher-dimensional models, where a similar
phenomenon may well be present.

\vspace{0.3cm}
\noindent {\it Acknowledgements.} We thank J.\ Ambj\o rn,
G.\ `t Hooft and S.\ Zohren for enjoyable discussions. Support through the
EU network on ``Discrete Random Geometry'', grant HPRN-CT-1999-00161,
is gratefully acknowledged.

 \end{document}
 \end